\documentclass[doublespacing]{elsart}

\begin{document}
\parindent0cm

\begin{frontmatter}

\title{Dynamic response of interacting one-dimensional fermions in the harmonic atom trap:
Phase response and the inhomogeneous mobility }

\date{\today}

\author{W. Wonneberger}

\ead{wolfgang.wonneberger@uni-ulm.de}

\address{Abteilung f\"ur Mathematische Physik, Universit\"at Ulm, D89069 Ulm, Germany}

\begin{abstract}
The problem of the Kohn mode in bosonized theories of one-dimensional interacting 
fermions in the harmonic trap is investigated and a suitable modification of the 
interaction is proposed which preserves the Kohn mode. The modified theory is used 
to calculate exactly the inhomogeneous linear mobility $\mu(z,z_0;\omega)$ at position $z$ 
in response to a spatial force pulse at position $z_0$. It is found that the inhomogeneous 
particle mobility exhibits resonances not only at the trap frequency $\omega _\ell $ 
but also at multiples $m \tilde{\epsilon}$, $m=2,3,...$ of a new renormalized collective 
mode frequency which depends on the strength of the interaction. In contrast, the homogeneous 
response obtained by an average over $z_0$ remains that of the non-interacting system. 
\end{abstract}


\end{frontmatter}

\section{Introduction}

Under suitable conditions, a gas of interacting fermions in one spatial dimension can be 
asymptotically described by a bosonic phase operator obeying a simple equation of motion. 
Conditions for this correspondence between fermionic and bosonic descriptions are a linear 
dispersion of free particle states allowing bosonization
via Kronig's identity \cite{deK35,Schoe04} and the addition of an anomalous vacuum which 
requires large fermion numbers not to cause harm. Furthermore the bosonic Hamiltonian should 
be bilinear in the field operator to facilitate diagonalization. The best known example is 
the Luttinger model \cite{T50,L63,ML65,Haldane}. One-dimensional systems are interesting 
because even weak interactions transform the quasi-particles of Fermi liquid theory into 
collective excitations of 
density wave type, and correlation functions are characterized by non-universal power laws
determined by one basic coupling constant $K$ (in case of one component, i.e. spin-polarized
fermions). Similar techniques can be applied directly to interacting Bose gases as has been 
done recently with ultracold quantum gases \cite{Mo98,Caz02,Caz03} based on pioneering 
work by Haldane \cite{Hal81a} and reviewed in \cite{Caz04}. Applications to mixtures of
one-dimensional bosons and fermions were given in \cite{CaHo04}.
Luttinger methods (for reviews cf \cite{E79,So79,Sch95,V95}) were extended to include
a trapping potential \cite{RZ03,KGH05} by means of the local density approximation. 
In \cite{Wo01,GW02,GGLW03,AGW04,GW04} attempts were made to include exactly
- as in the Calgero Sutherland model \cite{C69,S71} - a harmonic trapping potential into the 
Luttinger approach utilizing the linear dispersion
of free oscillator states, a concept recently adopted to quantum dissipation \cite{MaGo04}. 
In the case of interacting quantum gases the penalty are truncated interactions in order to render 
the model exactly solvable. They then are no more faithful representations of real interactions.
Furthermore, such type of approximation may violate essential symmetries of the original problem.
For harmonically trapped quantum gases Kohn's theorem \cite{Ko61,Brey89,Do94} is such a symmetry. 
It states that the response to a homogeneous external field with dipole coupling to the system 
displays a resonance exactly at the trap frequency irrespective of any translation invariant 
particle-particle interactions. This raises the question which many-body approximations 
respect the Kohn theorem. This problem was intensively studied in the case of ultracold quantum 
gases and mainly in the context of bosons. In \cite{SS96} it was found that the local density 
approximation at zero temperature is conserving. At finite temperature T the random phase 
approximations with exchange as well as the Hartree-Fock-Bogoliubov approximation are also 
conserving as argued in \cite{MT97}. 
The problem was taken up in \cite{FR98} and the Bogoliubov approximation was proven to be 
conserving at zero temperature. In \cite{BSt99} it was pointed out that the Gross-Pitaevskii 
equation is non-conserving at $T >0$ because the thermal cloud of the Bose-Einstein condensate 
is neglected. The work \cite{ZNG99} showed that kinetic equations extending the Hartree-Fock-Popov 
method are conserving. The general dielectric approach was proven in \cite{RBGS01} to be conserving 
too. Recently it was found \cite{O04} that superfluid fermions with a Feshbach resonance described 
by the generalized random phase approximation respect the Kohn theorem. 
The method of bosonization is intrinsically approximate and raises the same question.  
The phase formulation of a "Tomonaga-Luttinger model with harmonic confinement" was given 
in \cite{GW04}. In studying its linear response it turned out that the model in its original form 
violates Kohn's theorem. 
Here, we propose a method to remedy this deficiency. We then go on to calculate explicitely the 
inhomogeneous particle mobility in response to a spatially localized 
time varying force for a one-component gas. To this order we set up a linear but non-local 
differential equation for a phase operator adopted to the problem of harmonically trapped fermions. 
We solve it by means of its Green's function.

\section{The Kohn mode and bosonization}

In a strictly harmonic trap the trapping potential raises the translation mode at zero frequency 
to the trap frequency $\omega _\ell$ independent of inter-particle interactions provided these are
translation invariant. In a microwave experiment, when the electric field is dipole 
coupled to harmonically trapped charges, the resonance occurs exactly at this frequency. 
This was first noted by Kohn \cite{Ko61} in the context of magnetoresonance and later generalized
to solid state systems \cite{Brey89}. It was also discussed for ultracold quantum gases \cite{Do94}.
In the latter case of neutral atomic gases the effect is described in terms of a "sloshing mode":  
A homogeneous mechanical force alternating in time leads to a resonance exactly at the frequency 
$\omega _\ell$. A simple theoretical route to discuss the implications of the Kohn theorem is to 
consider the operator $\hat{z}_S$ of the center of mass (c.m.) position. In second quantization 
it is

\begin{eqnarray}\label{1}
\hat{z}_S= \frac{1}{N} \int^\infty _{-\infty}dz\, \hat{\psi}^+(z)z\,\hat{\psi}(z).
\end{eqnarray}

Transforming from local creation and annihilation operators to the harmonic oscillator
representation gives

\begin{eqnarray}\label{2}
\hat{z}_S= \frac{1}{N \alpha \sqrt{2}}\,\left( \sum ^\infty _{n=0}\sqrt{n+1} \,\hat{c}^+ _{n+1}
 \hat{c}_{n}+\sum ^\infty _{n=1}\sqrt{n} \,\hat{c}^+ _{n-1} \hat{c}_{n}\right).
\end{eqnarray}
Here, $\alpha$ is the inverse oscillator length and $N$ the fermion number.
For large particle numbers $N$, a condition implicit in the method of bosonization, the square
roots can be replaced by $\sqrt{N}$ \cite{MaGo04}. The bosonization prescription for density 
fluctuation operators $\hat{\rho}(p) \equiv \sum _{q} \hat{c}^+_{q+p} \hat{c}_{q}$
(cf. e.g. \cite{V95,Sch95}) according to

\begin{eqnarray}\label{4}
\hat{\rho} (p) = \left \{
\begin{array}{lll}
\sqrt{|p|} & \hat{d}_{|p|}, & p< 0,\\
\sqrt{p}   & \hat{d}^+_{p}, & p> 0,
\end{array} \right. 
\end{eqnarray}
where the operators $\hat{d}$ and $\hat{d}^+$ destroy and create collective bosonic 
excitations \cite{Haldane}, then leads to

\begin{eqnarray}\label{5}
 \hat{z}_S = \frac{1}{k_F}\left (\hat{d}_1 + \hat{d}^+ _1  \right) =
 \frac{2}{\pi k_F} \int ^0 _{-\pi}du \sin u \,\hat{\phi}_{odd}(u).
 \end{eqnarray} 
The Fermi wave number $k_F$ is given by $k_F=\alpha \sqrt{2 N}$.
The phase operator used to bosonize fermionic destruction and creation
operators in the formulation \cite{Schoe04} is

\begin{eqnarray}\label{6}
\hat{\phi}(u) = - i \sum ^\infty _{m=1} \frac{1}{\sqrt{m}}\, e^{i m (u+i \eta)}\,\hat{d}_{m },
\quad -\pi \le u \le \pi,
\end{eqnarray}
with a positive infinitesimal $\eta$. In \cite{AGW04} we found that the physical phase 
operator determining the particle density
is, however,  

\begin{eqnarray}\label{7}
\hat{\phi}_{odd}(u)&=&\frac{1}{2}\,(\hat{\phi}(u)+\hat{\phi}^+(u)-\hat{\phi}(-u)
-\hat{\phi}^+(-u))
\\[4mm]\nonumber
&\equiv& \sum _{n=1}^\infty \frac{1}{\sqrt{n}}\,
e^{-m \eta}\, \sin (n u) \left(\hat{d}_n + \hat{d} _n^+\right).
\end{eqnarray}

The Kohn mode requires a total Hamiltonian which gives the same Heisenberg equation 
of motion for the c.m. coordinate

\begin{eqnarray}\label{8}
 d^2_t \hat{z}_S(t) = - \frac{1}{\hbar ^2}\left [ \left [ \hat{z}_S, \hat{H}  \right], \hat{H} 
 \right]= - \omega ^2_\ell \hat{z}_S(t)
 \end{eqnarray}
as the free Hamiltonian

\begin{eqnarray}\label{9}
 \hat{H}_0=\frac{1}{2}\,\hbar\omega _\ell \sum _{m=0}^\infty
 (\hat{d}^+_m \hat{d}_m +\hat{d}_m \hat{d}^+_m).
 \end{eqnarray}

We have introduced earlier \cite{Wo01,GGLW03} a Luttinger-like model of interacting one-dimensional
fermions in the harmonic trap where the interaction is given by

\begin{eqnarray}\label{10}
 \hat{V}_c = - \frac{1}{2}\,V_c \sum ^\infty _{m=1}m \left (\hat{d}^2 _m +  \hat{d}^{2+}_m \right).
 \end{eqnarray}

This interaction approximates p-wave scattering in the one-component gas \cite{GlW04,AGW04}
in the simplest possible way. It is obvious that the condition (\ref{8}) is violated because 
the part

\begin{eqnarray}\label{11} 
 \hat{V}_{c1} \equiv -\frac{1}{2}\,V_c \left (\hat{d}^2 _1 + \hat{d}^{2+}_1 \right) 
 \equiv \frac{1}{2}\,\hbar \omega _\ell \tilde{V}_c \left (\hat{d}^2 _1 + \hat{d}^{2+}_1 \right)
 \end{eqnarray}
in the interaction results in $d^2_t \hat{z}_s(t) = - \tilde{\epsilon}^2 \hat{z}_s(t)$ with
 
 \begin{eqnarray}\label{12}
 \tilde{\epsilon}^2 = \omega ^2 _\ell \left (1-\tilde{V}^2 _c  \right) \neq \omega ^2 _\ell
 \end{eqnarray}
 for the interacting system. Quantum mechanically, the Kohn mode is a coherent state which would be
 turned into a squeezed state with renormalized frequency by the interaction equation (\ref{11}). 
 This is not admissible and one must subtract out that part of the interaction.
 We are mainly interested in the equation of motion for $\hat{\phi}_{odd}$ because it determines
 the response to external perturbations in the most direct way.
 One then must work out the consequences of the subtraction prescription for the phase formulation 
 of the theory. In the first step one introduces the phase operator form of $\hat{V}_{c1}$ using 
 
 \begin{eqnarray}\label{13}
 \hat{d}_1= \int ^\pi _{-\pi} \, du \,e^{iu}\left \{\frac{1}{2 \pi} \partial _u \hat{\phi}_{odd}(u) 
 + \hat{\Pi} (u)   \right\}.
 \end{eqnarray}
 
The momentum density $\hat{\Pi}$ is

\begin{eqnarray}\label{14} 
\hat{\Pi}(u) = - \frac{i}{2 \pi}\sum ^\infty _{n=1} \sqrt{n}\,e^{- n \eta} \sin n u \left (
\hat{d}_n - \hat{d}^+ _n \right), 
\end{eqnarray}
with the commutator

\begin{eqnarray}\label{15} 
[\hat{\phi}_{odd}(u),\hat{\Pi}(v)]=\frac{i}{2}\,\delta (u-v)
\end{eqnarray}
in the reduced interval  $I_\pi \equiv (-\pi < u < 0)$. This shows that $2\hbar\hat{\phi}_{odd}$ 
and $\hat{\Pi}$ are canonically conjugate on this interval.
In applications the auxiliary variable $u$ is related to the physical position $z$ inside 
the harmonic trap by 

\begin{eqnarray}\label{17}
u \rightarrow u_0(z)= \arcsin\left(\frac{z}{L_F}\right)-\frac{\pi}{2},\quad z=L_F \cos(u_0), 
\end{eqnarray}
where $L_F=\sqrt{\hbar 2N}/\alpha $ is half the quasi-classical extension of the Fermi sea.  
The non-linear relation between the formal variable $u$ and physical position $z$ 
expresses the trap topology. It is also obvious that the reduced interval $I_\pi$
is sufficient to describe the physics.

The incriminating part in the original interaction becomes
 
 \begin{eqnarray}\label{16}
 \hat{V}_{c1} &=& - V _c \int ^\pi _{-\pi} \, du du' \,e^{iu+iu'}
  %\\ \nonumber
 \left \{ \frac{1}{4 \pi ^2} \partial _u \hat{\phi} _{odd}(u)\partial _{u'}\hat{\phi}_{odd}
 (u') + \hat{\Pi} (u)\hat{\Pi}(u')   \right\}
 \\[4mm] \nonumber
 &=& \hbar \omega _\ell \tilde{V}_c \int ^0 _{-\pi} \, du du' \left \{4\sin u \sin u' \hat{\Pi} (u)
 \hat{\Pi}(u') - \frac{1}{\pi ^2} \cos u \cos u'\partial _u \hat{\phi} _{odd}(u)\partial _{u'}
 \hat{\phi}_{odd} \right\}.
 \end{eqnarray}
 
The phase form of the reduced Hamiltonian (neglecting the zero-mode which would cancel in subsequent
calculations) is
 
 \begin{eqnarray}\label{18}
 \tilde{H}_{red}= \frac{\epsilon}{2} \int ^\pi _{-\pi} \, du \left [ 2 \pi K \,\hat{\Pi}^2 (u)+ 
 \frac{1}{2 \pi K} \left ( \partial _u \hat{\phi}_{odd} (u) \right) ^2   \right]-\hat{V}_{c1},
 \end{eqnarray}
 with

 \begin{eqnarray}\label{19}
 K=\sqrt{\frac{1+\tilde{V} _c }{1-\tilde{V}_c }},\quad\epsilon =\hbar \tilde{\epsilon} 
 = \hbar\omega _\ell\frac{2K}{K^2+1} \equiv \hbar \omega _\ell \sqrt{1-\tilde{V}_c^2}.
 \end{eqnarray}
 
 The first part of the reduced Hamiltonian is diagonal with renormalized excitation energies 
 $m \epsilon$, $m=1,2,3...$, while the second part is a non-local but still bilinear correction.
 The same results can be obtained by a selective renormalization of the trap frequency 
 similar to the well known procedure in coupling a linear bath to a harmonic oscillator:
  
\begin{eqnarray}\label{20}
 \hat{H}_0 \rightarrow \hat{H}_0+ \frac{1}{2}\,\hbar(\tilde{\omega}_\ell-\omega _\ell) 
 (\hat{d}^+_1 \hat{d}_1 +\hat{d}_1 \hat{d}^+_1),\quad \tilde{\omega}_\ell\equiv \omega _\ell 
 \sqrt{1+\tilde{V}^2 _c}. 
 \end{eqnarray}    
 
\section{Inhomogeneous linear mobility}

The inhomogeneous mobility is the particle current response at position $z$ to a $\delta$-function 
force $f$ 

\begin{eqnarray}\label{21}
f(z,z_0;t)= F(t)\,\delta(z-z_0)
\end{eqnarray}
at a different position $z_0$ inside the trap. Quantitatively, the inhomogeneous linear 
mobility $\mu(z,z_0;\omega)$ relates the expectation value of the local particle current 
density operator $\hat{j}_p(z)$ to the force by

\begin{eqnarray}\label{22}
 \langle \hat{j}_p(z) \rangle _\omega&=&\int _{-\infty}^\infty dt 
\,e^{i \omega t}\, \langle \hat{j}_p(z)
\rangle _t 
\\[4mm]\nonumber
&=&\int dz'\,\mu(z,z';\omega)\,f_\omega(z',z_0)=\mu(z,z_0;\omega)\,F_\omega.
\end{eqnarray}

The Hamiltonian for a time-dependent, spatially localized external force $f(z,z_0;t)$
affecting N fermions on the z-axis is

\begin{eqnarray}\label{23}
 \hat{H}_{ext} (z_0,t) = - F(t)\,\sum _{n=1}^N \Theta(\hat{z}_n-z_0).  
\end{eqnarray}

In second quantized form it becomes

\begin{eqnarray}\label{24}
 \hat{H}_{ext} (z_0,t) = - F(t)\,  \int _{z_0}^\infty dz\, \hat{\psi} ^+ (z) \hat{\psi}(z). 
\end{eqnarray}

For simplicity, a one-component gas of spin polarized fermions is assumed. Without enhancement 
of the interactions by Feshbach resonances \cite{R03} this is an admittedly academic case 
when regarding the role of interactions.
The total particle density operator contains a part $\delta\hat{\rho}_{slow}$ which varies 
slowly in space. In an appropriate WKB  expansion (for large $N$), it is given by \cite{AGW04}  

\begin{eqnarray}\label{25}
\delta\hat{\rho}_{slow}(z)= \frac{1}{\pi}\,\partial _z \hat{\phi}_{odd}(u_0(z)).
 \end{eqnarray}

Again this has an analogous expression in the theory of Luttinger liquids and its bosonic
equivalent \cite{Hal81a,Caz04}.
However, in the present case, there is a non-linear relation between the argument $u_0$ of the 
phase operator and the actual spatial position $z$ according to equation (\ref{17}). There
is also a rapidly varying Friedel part in the particle density which results from the confinement. 
Away from the classical boundaries $z=\pm L_F$ it gives a small
correction to the mobility if we average the latter over a small length $L$ around $z_0$ with 
$L_F \gg L \gg 1 /k_F$. It will be neglected here. Utilizing the continuity equation leads to 

\begin{eqnarray}\label{26}
\partial _t \delta\hat{\rho}_{slow}(z,t)&=&\frac{1}{\pi}\,\partial _z \left(\partial _t 
\hat{\phi}_{odd}(u_0(z),t)\right)
= -\partial _z \,\hat{j}_{slow}(z,t),
\end{eqnarray}
and shows that the slow  current density is given by

\begin{eqnarray}\label{27}
 \hat{j}_{slow}(z,t)= -\frac{1}{\pi}\,\partial _t \hat{\phi}_{odd}(u_0(z),t).
 \end{eqnarray}

We make the identification $\hat{j}_p \equiv\hat{j}_{slow}$ which according to equation 
(\ref{22}) leads to the generic form 

\begin{eqnarray}\label{27a}
 \mu(\omega)=\frac{i \omega \langle \hat{\phi}_{odd}\rangle _\omega}{\pi F_\omega}.
 \end{eqnarray}
for the mobility. Coupling the external force to the slowly varying density amounts to the 
replacement 

\begin{eqnarray}\label{28} 
\hat{\psi}^+ (z) \hat{\psi} (z)  \rightarrow \delta \hat{\rho}_{slow}(z)=\frac{1}{\pi}\, 
\partial _z\hat{\phi}_{odd}(u_0(z)) 
\end{eqnarray}
in equation (\ref{24}) and leads to the external phase Hamiltonian

\begin{eqnarray}\label{29}
\hat{H}_{ext} = F(t)\,\frac{1}{\pi}\,\hat{\phi}_{odd}(u_0(z_0)).
\end{eqnarray}
 
\section{Equations of motion}

The equations of motion are most easily obtained via functional derivatives using
 
 \begin{eqnarray}\label{30}
 2 \hbar \partial _t \hat{\phi}_{odd}(u,t) = \frac{\delta \hat{H}_{red}}{\delta \hat{\Pi}(u,t)},
 \quad
% \\[4mm]\nonumber
 2 \hbar \partial _t \hat{\Pi}(u,t) = -\frac{\delta \hat{H}_{red}}{\delta \hat{\phi}_{odd}(u,t)}.
 \end{eqnarray}
 
 One obtains:
 
 \begin{eqnarray}\label{32}
 \partial _t \hat{\phi}_{odd} = 2 \pi \tilde{\epsilon}\, \hat{\Pi}
 %\\[4mm]\nonumber
 - 4 \omega _\ell \tilde{V}_c \,\sin u \int ^0_{-\pi}du' \sin u'\,\hat{\Pi}(u')
  \end{eqnarray}
 and
 
 \begin{eqnarray}\label{33}
 \partial _t \hat{\Pi} &=& \frac{\tilde{\epsilon}}{2 \pi K}\partial^2 _u \hat{\phi}_{odd}
 \\[4mm]\nonumber
 &-&\frac{1}{\pi^2}\, \omega _\ell \tilde{V}_c \,\sin u \int ^0_{-\pi}du' \sin u'\,\hat{\phi}_{odd}(u')
  %\\[4mm]\nonumber
  -F(t)\,\frac{1}{2 \pi \hbar}\,\delta(u_0(z_0)-u).
  \end{eqnarray}

The equation of motion for $\hat{\phi}_{odd}$ closes at the level of the second time derivative:
 
 \begin{eqnarray}\label{34}
 \partial _t ^2 \hat{\phi}_{odd} &=& \tilde{\epsilon}^2 \partial _u ^2 \hat{\phi}_{odd}  
-  \omega _\ell ^2 \tilde{V}_c ^2 \sin u \left[\frac{2}{\pi}\int ^0 _{-\pi} \, du' \,\sin u'
\hat{\phi} _{odd}(u')\right]
\\[4mm]\nonumber
&-&F(t)\frac{\tilde{\epsilon}K}{\hbar}\,\delta(u_0(z_0)-u)+F(t)\frac{2 \omega _\ell 
\tilde{V}_c}{\pi \hbar}\sin u_0(z_0) \sin u.
\end{eqnarray}
 Here, essential use has been made of the relation
  
 \begin{eqnarray}\label{35}
 \tilde{\epsilon}\left(K-\frac{1}{K}\right)=2 \omega _\ell \tilde{V}_c.
 \end{eqnarray}

 The non-local operator results from the subtraction procedure and projects onto the subspace of 
 the Kohn mode:
 This is best seen by considering excitations $\varphi (u,t) = \langle \hat{\phi}_{odd}(u)
 \rangle _t$ i.e. averaging the equation of motion over a non-equilibrium state. For $F=0$ one gets 
 
 \begin{eqnarray}\label{36}
 \partial _t ^2 \varphi = \tilde{\epsilon} ^2 \partial _u ^2 \varphi - \omega ^2 _\ell \tilde{V}_c^2
  \sin u \left[\frac{2}{\pi}\int ^0 _{-\pi} \, du ' \sin u ' \,\varphi (u',t)\right].
 \end{eqnarray}
 
 The Kohn mode is a density (dipole) oscillation with $\varphi _K \propto \sin u_0(z)$ and associated
 real space density modulation $\delta \rho(z) \propto \partial _z \varphi _K \propto z/\sqrt{1-z^2/L_F^2}$.
 It solves equation (\ref{36}): 
 $\partial _t ^2 \varphi _K = - \tilde{\epsilon} ^2 \varphi _K - \omega ^2 _\ell \tilde{V}^2 _c 
 \varphi _K$ with the correct frequency because of  $\tilde{\epsilon} ^2 = \omega ^2 _\ell 
 (1-\tilde{V}_c^2)$.
 The non-local projection operator has no influence on modes orthogonal to the Kohn mode.
 Their excitation energies scale with $\tilde{\epsilon}$, the renormalized frequency of the model.

\section{Calculation of inhomogeneous linear mobility}

 The eigenvalue problem associated with the phase equation
 
 \begin{eqnarray}\label{38}
 - \omega ^2 \varphi _\omega = \tilde{\epsilon} ^2 \partial _u ^2 \varphi _\omega - \omega ^2_\ell
 \tilde{V}_c^2 \sin u \left[\frac{2}{\pi}\int ^0 _{-\pi} \, du ' \sin u ' \,\varphi _\omega(u')\right]
  \end{eqnarray}
  can be cast into a convenient form by introducing the function
 
 \begin{eqnarray}\label{39}
 \varphi _1 (u)\equiv \sqrt{\frac{2}{\pi}} \sin u,
 \end{eqnarray}
 which is normalized in $I_\pi= (-\pi,0)$ and satisfies the boundary conditions implied by $
 \hat{\phi}_{odd}(u)$, namely $\varphi _1 (-\pi) = 0= \varphi _1 (0)$. Introducing the linear 
 symmetric operator ${\bf L}_\omega$
  
 \begin{eqnarray}\label{40}
 {\bf{L}} _\omega \equiv \frac{\tilde{\epsilon}^2}{\omega _\ell ^2} \,\partial ^2 _u +
 \frac{\omega ^2}{\omega ^2 _\ell}
 - \tilde{V}_c ^2 \,\varphi _1 (u) \int^0 _{-\pi} \, du ' \varphi _1 (u')\,(....),
 \end{eqnarray}
 the eigenvalue problem becomes
 
 \begin{eqnarray}\label{41}
 {\bf{L}} _\omega \varphi _n = - \lambda _n^2 (\omega)\varphi _n.
 \end{eqnarray}
  
 Because of the relation $\tilde{\epsilon}^2 \equiv \omega _\ell ^2 \left (1-\tilde{V}_c ^2  
 \right)$ it is easy to see that $\varphi _1 (u)$ is an eigenfunction with eigenvalue
 
 \begin{eqnarray}\label{42}
 \lambda _1 ^2(\omega) = 1 -\frac{\omega ^2}{\omega _\ell^2}.
 \end{eqnarray}
 
 Furthermore,
 
 \begin{eqnarray}\label{43}
 \varphi _n (u) = \sqrt{\frac{2}{\pi}}\, \sin n u, \quad n=1,2,...
 \end{eqnarray}
 is an orthonormal basis on $I_\pi$. It is thus clear that for $n \ge 2$
 
 \begin{eqnarray}\label{44}
 {\bf{L}} _\omega \varphi _n = \left (-n ^2 \frac{\tilde{\epsilon} ^2}{\omega _\ell ^2} 
 + \frac{\omega ^2}{ \omega _\ell ^2}
 \right) \varphi _n \equiv - \lambda _n^2 (\omega)\, \varphi _n,
 \quad \lambda _n ^2(\omega) &=& n ^2 \frac{\tilde{\epsilon}^2}{\omega _\ell ^2}
 - \frac{\omega ^2 }{\omega _\ell ^2}
 \end{eqnarray}
 holds. The relevant Green's function is 
 
 \begin{eqnarray}\label{45}
 G_\omega (u,u') = - \sum ^\infty _{n=1} \frac{\varphi _n (u)\varphi _n (u')}{\lambda^2 _n(\omega)}.
 \end{eqnarray}
 
This allows to solve for the inhomogeneous local mobility according to equation (\ref{27a})
with $\langle \hat{\phi}_{odd} \rangle _\omega \equiv \varphi _\omega(u_0(z),u_0(z_0))$.
Utilizing the identity $\tilde{\epsilon}K  \equiv \omega _\ell(1+\tilde{V}_c)$ leads to

\begin{eqnarray}\label{48}
\varphi _\omega = - \frac{F _\omega \omega _\ell}{\hbar } \frac{\varphi _1 (u_0(z) )\varphi _1 (u_0(z_0))}
{\omega^2 _\ell -\omega ^2}
- \frac{F _\omega \tilde{\epsilon}K}{\hbar } \sum ^\infty
_{n=2} \frac{\varphi _n (u_0(z)) \varphi _n (u_0(z_0))}{n^2 \tilde{\epsilon}^2-\omega ^2}.
\end{eqnarray}

Using $\sin u_0(z)= -\sqrt{1-z^2/L_F^2} \equiv -Z(z)$ finally gives

\begin{eqnarray}\label{49}
\mu (z, z_0;\omega) = - \frac{2 i \omega \omega _\ell}{\pi^2 \hbar} 
\frac{Z(z) Z(z_0)}{\omega _\ell^2- \omega ^2}
- \frac{2 i \omega \tilde{\epsilon}K}{\pi^2 \hbar} \sum ^\infty _{n=2}
\frac{\sin n u_0 (z)\sin n u_0(z_0)}{n^2 \tilde{\epsilon}^2- \omega ^2}.
\end{eqnarray}
Analyticity of the response function for $\Im \omega \ge 0$ requires $\omega \rightarrow \omega 
+ i \eta$ with a positive infinitesimal $\eta$.
It is seen that the inhomogeneous mobility has resonances at all excitation frequencies including 
$\omega _\ell$. This applies also to the non-interacting case. It is only the homogeneous response
which is trivialized by Kohn's theorem. In order to see this, we calculate the homogeneous 
mobility via

\begin{eqnarray}\label{50}
\mu (z, \omega) = - L_F \int ^0_{-\pi}du_0 \sin u_0 \,\mu(z,z_0(u_0);\omega),
\end{eqnarray}
and find 

\begin{eqnarray}\label{51}
\mu (z, \omega) = \frac{L_F}{\pi \hbar}\,\frac{i  \omega \omega _\ell}{\omega ^2 - \omega ^2 _\ell}\, Z(z).
\end{eqnarray}
This is identical to the local response of non-interacting fermions in the harmonic trap 
to a homogeneous force. The Kohn theorem is clearly fulfilled.

Returning to the inhomogeneous mobility, the sum can be evaluated analytically. By including 
the $n=1$ term in the summation one obtains

\begin{eqnarray}\label{52}
\mu (z,z_0; \omega) &=& -\frac{i K}{2 \pi \hbar \sin(\pi a)}
\\[4mm]\nonumber
&&\times\{\cos\left[a(\pi +u_0(z)+u_0(z_0))\right]-\cos\left[a (\pi-|u_0(z)-u_0(z_0)|)\right]\}
\\[4mm]\nonumber
&+&\frac{2  \tilde{V}_c \,Z(z) Z(z_0)}{\pi^2  \hbar}\left\{\frac{i \omega \omega _\ell((1+\tilde{V}_c)
\omega ^2_\ell-\omega^2)}{(\omega^2 _\ell-\omega^2)((1-\tilde{V}_c^2)\omega _\ell^2-\omega ^2)}\right\}.
\end{eqnarray}
with $a=\omega/\tilde{\epsilon}$. Here, the first term also includes a contribution from the 
Kohn mode $\varphi _1$ which removes all renormalizations due to the interaction from the second term 
in going over to the homogeneous response. By this mechanism Kohn's theorem is restored.

 \section*{Acknowledgments}

The author thanks S. N. Artemenko and F. Gleisberg for helpful discussions and Deutsche 
Forschunsgemeinschaft for financial help.

\end{document}